\documentclass[twocolumn,showpacs,preprintnumbers,amsmath,amssymb]{revtex4}

\usepackage[english]{babel}
\usepackage[T1]{fontenc}
\usepackage{textcomp, comment}
\usepackage[utf8]{inputenc}
\usepackage{amsmath, amssymb}
\usepackage{graphicx}
\usepackage{dcolumn}

\DeclareMathOperator{\erfc}{erfc}
\newcommand{\dfnd}{:=}
\newcommand{\D}{D}

\newcommand{\dx}{
  \left| \Delta \vec{x}_{\bot} \right|^2 
}

\newcommand{\diff}{\,\mathrm{d}}

\begin{document}

\preprint{APS/123-QED}

\title{Numerical study of the Transverse Diffusion coefficient \\
  for a one component model of a plasma}

\author{ Lorenzo Valvo}
\email{lorenzo.valvo@studenti.unimi.it}
\affiliation{Corso di laurea in
    Fisica, Università degli Studi di Milano, Via Celoria~16, 20133
    Milano, Italy}
\author{Andrea Carati}
 \email{carati@mat.unimi.it}
\affiliation{%
Department of Mathematics, Universit\`a degli Studi di Milano,
Via Saldini 50, 20133 Milano, Italy
}%

\date{\today}

\begin{abstract}
We report the results of MD numerical simulations for a one component
model of a plasma in the weakly coupled regime, at different values of
temperature $T$ and applied magnetic field $\vec B$, in which the
diffusion coefficient $D_{\perp}$ transverse to the field is
estimated.  We find that there exists a threshold in temperature, at which
an inversion occurs, namely, for $T$ above the threshold the diffusion
coefficient $D_{\perp}$ starts decreasing as $T$ increases. This is at
variance with the behavior predicted by the Bohm law $D_{\perp}\sim
T/B$, which actually holds below the threshold. In addition we find that, for
temperatures above such a threshold, another transition occurs, now
with respect to the values of the magnetic field: for weak magnetic fields the
diffusion coefficients scales as $1/B^2$, in 
agreement with the predictions of the standard kinetics theory, while it
apparently saturates when the field strength is sufficiently increased.
\end{abstract}


\maketitle


In this letter we report the results of some Molecular Dynamics
simulations of a one component model of plasma. In particular we 
estimate the diffusion coefficient $D_{\perp}$ transverse to the
magnetic field, in the case of a weakly coupled plasma, for different
values of temperature $T$ and of magnetic field strength $B$.

Up to now, Molecular Dynamics simulations for the diffusion
coefficient were implemented only for the case of strongly coupled
plasmas, see for example the recent paper~\cite{OttBonitz}. The
general conclusion of that paper can be summarized by saying that, in
the strongly coupled case, the diffusion coefficient
obeys the scaling law $D_{\perp}\propto T B^{-1}$ 
proposed long ago by Bohm
(see Table~1, page 135003-2 of \cite{OttBonitz}).

Our computations, while confirming the ones of paper~\cite{OttBonitz} for
the strongly coupled case, indicate that some important differences arise in the
weakly coupled regime. In particular there exists a threshold in the
coupling parameter, such that 
\begin{itemize}
  \item concerning the dependence on $T$, at variance with  Bohm's
    law, above the threshold the diffusion coefficient $D_{\perp}$
    starts  decreasing as temperature increases;
  \item concerning the dependence on $B$, one finds that $D_{\perp}$ decreases
    as $B^{-2}$ for small fields, but then apparently saturates to a
    constant value for larger values of $B$.
\end{itemize}

\begin{figure}
  \centering
  \includegraphics[width=0.5\textwidth]{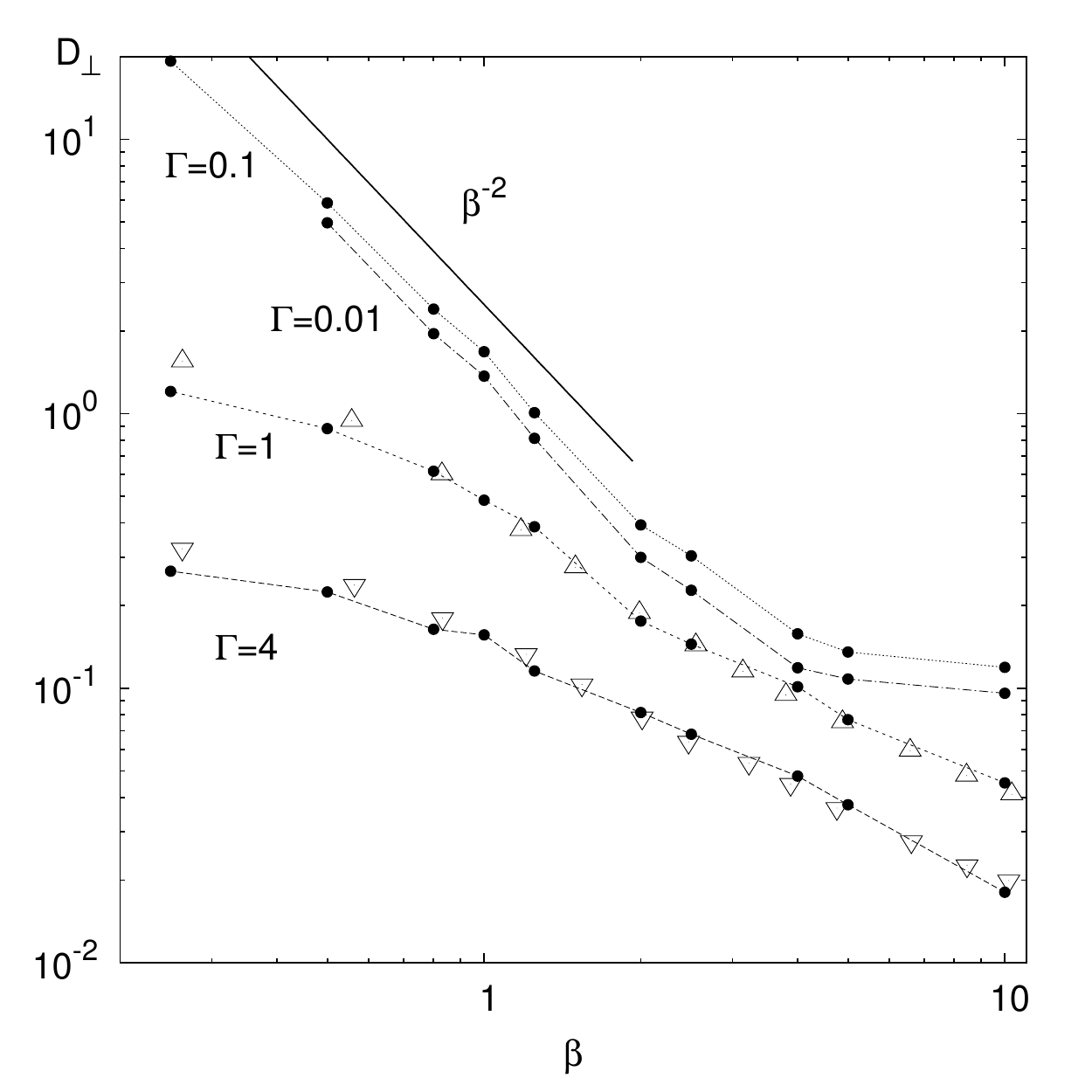}
  \caption{ Diffusion coefficient transverse to the magnetic field versus 
   $\beta$, as computed by MD simulations. Full circles are the results
    of our computations, while the empty triangles are values taken
    from paper~\cite{OttBonitz}. The straight line corresponding
    to $\beta^{-2}$ is also shown (dashed line).
   }\label{fig:graph_new}
\end{figure}

The results of our computations are summarized in
figure~\ref{fig:graph_new}, where, in logarithmic scale, the value of
the coefficient $D_{\perp}$ is reported (full circles) as a function
of the dimensionless parameter $\beta$, defined by $\beta =
B/\sqrt{nmc^2}$, where $m$ is the electron mass, $n$ the electron
density and $c$ the speed of light (we are working in the
c.g.s. system).  The different lines connect simulations performed at
the same value of $\Gamma$, the dimensionless coupling parameter
defined by $\Gamma = n^{1/3}e^2/(k_BT)$, where $k_B$ is the Boltzmann
constant, and $e$ the electron charge. Such parameter discriminates
between the strongly coupled case, corresponding to $\Gamma>1$, and
the weakly coupled case corresponding to $\Gamma \ll 1$.

For comparison, in the same figure are reported also some values
(empty triangles) taken from reference \cite{OttBonitz}, for $\Gamma$
equal to $1.24$ and $3.1$ (corresponding to the values $2$ and $5$
if one uses the definition of $\Gamma$ given in that paper). The
agreement with our data seems to be good.

The figure clearly exhibits that, while for large values of $\Gamma$,
actually up to $0.1$, the 
coefficient $D_{\perp}$ decreases as a function of $\Gamma$
(i.e. is an \textbf{increasing} function of $T$), for the smaller value
$\Gamma=0.01$ an inversion occurs, i.e. the values of $D_{\perp}$ are
\textbf{smaller} than the corresponding values at $\Gamma=0.1$. So
there must exist a threshold in $\Gamma$, below which 
$D_{\perp}$ becomes a \textbf{decreasing} function of $T$.

A straight line corresponding to $\beta^{-2}$ is also shown. One can
check that up to $\beta=1$, the data for $\Gamma=0.1$ and $\Gamma=0.01$
seem to lie parallel to such a line. This means that, at fixed
density, the coefficient $D_{\perp}$ decreases as $B^{-2}$, as
predicted by kinetic theory (see~\cite{Spitzer,Rosenbluth}). 
However, by further increasing the magnetic field above $\beta=1$, the
diffusion coefficient appears to saturate to an apparently constant
value independent of $B$. To our knowledge, this phenomenon was neither observed
nor foreseen before.

The only reported evidence of some kind of transition that should
occur in a weakly coupled plasma, when passing from $\beta \lesssim 1$
(weakly magnetized) to $\beta\gtrsim 1$ (strongly magnetized) was
given in~\cite{CBMZG}. In that paper, such a transition was ascribed
to a transition from a fully chaotic regime (low magnetic field) to a
partially ordered one (high magnetic field), as first proposed in
paper \cite{CZMMMG}. 

\begin{figure}
  \centering
  \includegraphics[width=0.5\textwidth]{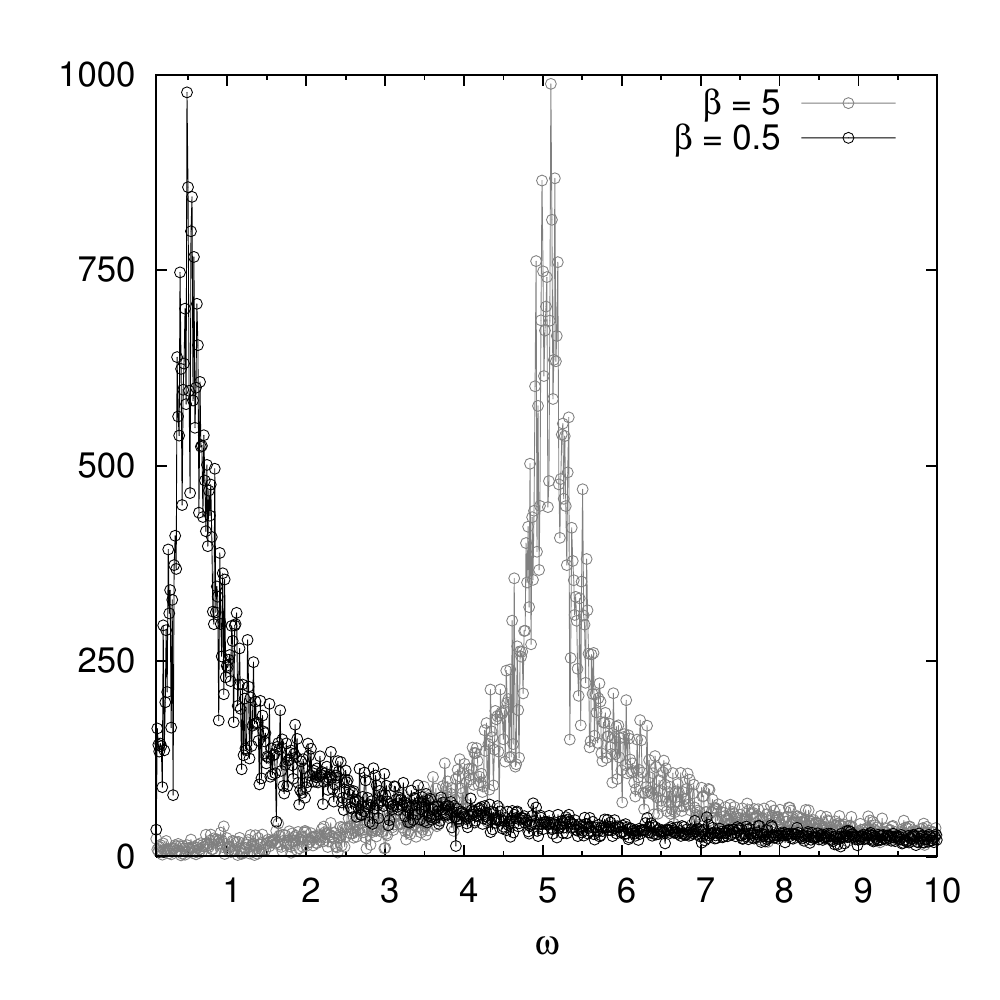}
  \caption{ Fourier transform of the velocity autocorrelation function
    $\langle \vec{v}_{\perp}(t)\cdot\vec{v}_{\perp}(0)\rangle$,
    obtained in two simulations, with the same value $\Gamma=0.1$, and
    two different values of $\beta$, namely, $\beta=0.5$ (black line)
    and $\beta=5.$ (gray line). The frequencies are reported in units
    of $\omega_p$. Notice the peaks centered at the corresponding
    cyclotron frequencies.
   }\label{fig:corr_vel}
\end{figure}

We now illustrate how a change of the dynamical behavior, might also
explain the behavior of $D_{\perp}$ reported above.  We recall (see
for example the textbook~\cite{Wannier}) that the diffusion
coefficient can be expressed in terms of the velocity autocorrelation
$\langle \vec {v}_{\perp}(t) \cdot \vec{v}_{\perp} (0)\rangle$ as follows
\begin{equation}\label{eq:diffusione}
  D_{\perp} = \frac 12 \int_0^{+\infty} \langle \vec {v}_{\perp}(t)
  \cdot \vec{v}_{\perp} (0)\rangle \diff t \ ,
\end{equation}
where $\vec{v}_{\perp}$ in the component of the velocity of a particle,
transverse to the magnetic field, and the brackets mean a suitable
average over the particles. Now,
in figure~\ref{fig:corr_vel} we
report the velocity spectrum, i.e. the Fourier transform of the
autocorrelation, calculated at $\Gamma=0.1$, 
for two values of $\beta$ (below and above 
$\beta=1.$). One sees that in both cases, a strong peak occurs at
the corresponding cyclotron frequency $\omega_c \dfnd eB/mc$, 
so that one can suppose that
$$
\langle \vec {v}_{\perp}(t) \cdot \vec{v}_{\perp} (0)\rangle \simeq
\langle \vec {v}_{\perp}(0) \cdot \vec{v}_{\perp} (0)\rangle
\cos(\omega_c) f(t) \ ,
$$
where $f(t)$ is a function which characterizes the decay to zero of the
autocorrelation as $t\to +\infty$.  In the fully chaotic regime one
can take $f(t)=e^{-\gamma t}$, where $\gamma$ is the inverse of the
decorrelation time. If such a time is larger than the cyclotron
period, from eq. (\ref{eq:diffusione}) one gets
\begin{equation}\label{eq:coeff1}
  D_{\perp} = \frac {k_BT}{m} \frac {\gamma}{\omega_c^2} \ .
\end{equation}
This expression shows that $D_{\perp}$ decreases as $B^{-2}$, in
agreement with our numerical data for small $B$.

Instead one can suppose that, in a partially ordered case, the decay of
correlations is much slower, for example as an inverse
power of time. Taking for example $f(t)= 1/\big(1+(\gamma^*
t)^2\big)$, one gets the expression
\begin{equation}\label{eq:coeff2}
  D_{\perp} = \frac {k_BT}{m} \frac{\pi}2 \frac
  {\exp(-\omega_c/\gamma^*)}{\gamma^*} \ . 
\end{equation} 
Now one needs to match the expression (\ref{eq:coeff1}) with
(\ref{eq:coeff2}) at $\omega_c$ equal to the plasma frequency
$\omega_p \dfnd \sqrt{e^2/n}$, i.e., for $\beta=1$, which is precisely the value
of the threshold predicted in reference~\cite{CZMMMG}. Given a
value of $\gamma/\omega_p$, this matching determines two possible values
of $\gamma^*/\omega_p$, one smaller than 1 and one bigger.
If one chooses the larger one, the  expression (\ref{eq:coeff2})
for $\omega_c\simeq\omega_p$ gives a curve with a slope much smaller
than the curve (\ref{eq:coeff1}), thus reproducing
the behavior found numerically. At very large values of $\omega_c$,
the diffusion coefficient $D_{\perp}$ should
begin to decrease faster than any inverse power of $B$, but this range
is actually outside our reach. At any rate, a decrease of $D_{\perp}$
faster than any inverse power of $B$, is reported in reference
\cite{Ranganathan}, and could actually be ascribed to the phenomenon 
just described.

\begin{figure}
  \centering
  \includegraphics[width=0.5\textwidth]{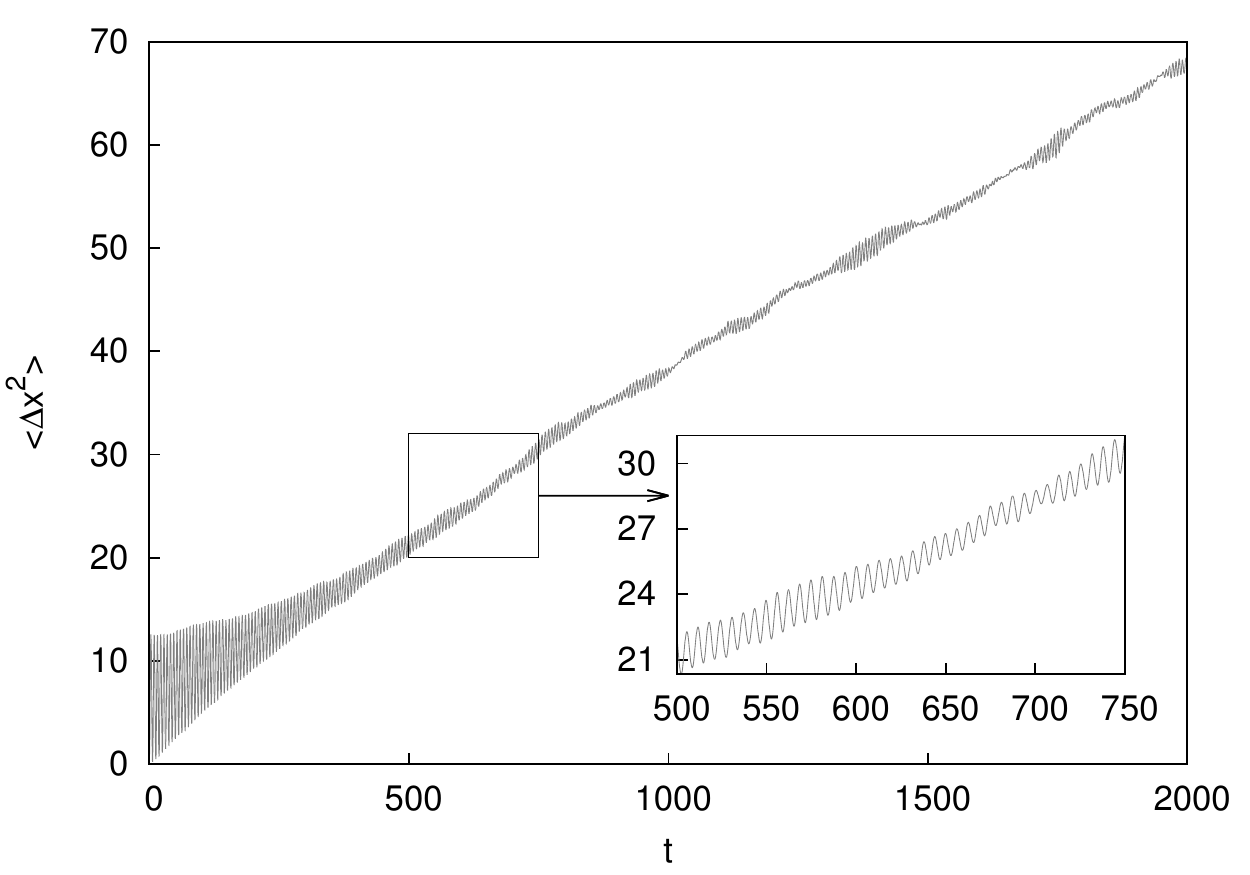}
  \caption{
    Graph of $\langle \dx \rangle$ vs time $t$ (in $\omega_c^{-1}$ units) 
     in the case $\Gamma=0.01$ and $\beta=4$. 
  }\label{fig:oscillations}
\end{figure}

We give now some details of our numerical computations.  First of
all, for the purpose of estimating $D_{\perp}$, it is better to use
directly its definition, instead of making use of relation
(\ref{eq:diffusione}).  We recall that the transverse diffusion
coefficient is defined by
\begin{equation}
  \D_{\perp} = \lim_{t\rightarrow\infty} \frac{\langle\dx\rangle}{4t} 
  \label{diffusionCoefficientDef}
\end{equation}
where 
\begin{equation}
  \dx \dfnd |x(t) - x(t_0)|^2 + |y(t) - y(t_0)|^2 \ ,
\end{equation}
i.e, as the mean square displacement of a particle in the plane
orthogonal to the magnetic field $\vec B$ (which is taken directed as
the $z$--axis) and the average should in principle be an average over
all the plasma particles.  The quantity in
eq.~\eqref{diffusionCoefficientDef} can be computed from a numerical
simulation by averaging over the particles which participate to the
simulation, while the asymptotic value can be found from the plot of
$\langle\dx\rangle$ versus $t$. A typical example of such a
plot is shown in Fig.~\ref{fig:oscillations}; as one can see, it
displays an oscillatory behavior superimposed to a linear growth, and
a ballistic motion at the beginning.  To remove the oscillations we
replaced each value of $\langle\dx\rangle(t)$ by its mean value
$\bar{X}(t)$ taken over a cyclotron period centered at $t$.  The
resulting data were analyzed with a linear regression of the law
$\bar{X} = \D t + C$, with two constants $C$ and $D$.  Here one has to
choose a temporal window to include only the tail of the graph; we
progressively restricted such window until the reduced error $\chi$
(the sum of the squared residual divided by the number of points minus
2) became less than 1.  The values thus found are reported in
figure~\ref{fig:graph_new}.

For what concerns the model, we recall that the one component model of a
plasma consists of a gas of electrons moving in a fixed uniform
neutralizing background. So we consider a number $N$ of electrons in a
cubic box of side $L$ with periodic boundary conditions, the electrons
being subject to mutual Coulomb interactions, and to an external
magnetic field $\vec{B} = B\vec{e}_z$. The density is then defined by
$n=N/L^3$.

If $t$ denotes time and $\vec{x}_i$ the position of the $i$-th electron
(with $i=1,\dots,N$), with the rescaling 
\begin{equation}
  \vec{y}_i = n^{-1/3}\vec{x}_i,\quad \tau = \omega_c t,\quad m=1 \ ,
  \label{units}
\end{equation}
which in particular implies that the density takes the value 1,
the equations of motion read 
\begin{equation}
  \label{newtonEquation}
  \frac{d^2 {\vec{y}}_i}{d\tau^2} = \vec{e}_z \times \frac{{d\vec{y}}_i}{d\tau} 
    +\frac{1}{\beta^2} \sum_{j \neq i} \vec{E}_j(\vec{y}_i)
\end{equation}
where $\vec{E}_j$ is the electric field created by the $j$--th electron,
evaluated at the position of the
$i$-th one. The total electric field $\vec{E}=\sum_{j \neq i}
\vec{E}_j$ acting on an electron, created by a periodic
system of charges, can be computed via the Ewald formula 
(see~\cite{Gibbon}), as follows
\begin{equation}
  \begin{split}
    &\vec{E}(\vec{y}_i) = \\  
      & \sum_{\vec{l}} \sum_{j=1}^N 
        \frac{ \vec{y}_{ij\vec{l}} }{| \vec{y}_{ij\vec{l}} |^3} 
	\left[ \erfc({\alpha} | \vec{y}_{ij\vec{l}} | ) 
        + \frac{2 \alpha | \vec{y}_{ij\vec{l}} | }{ \sqrt{\pi} }
        \exp(-{\alpha}^2{ | \vec{y}_{ij\vec{l}} |}^2) \right] \\
      &  + \frac{4 \pi}{N} \sum_{\vec{k} \neq 0} \sum_{j=1}^N
        \frac{\vec{k}}{k^2} e^{-k^2\!/4 {\alpha}^2} 
        \sin( \vec{k} \cdot \vec{y}_{ji}), 
	\qquad \alpha = \frac{\sqrt{\pi}N^{1/6}}{L} \notag
	\label{ewaldField} \\
  \end{split}
\end{equation}
Here $\vec{y}_{ij\vec{l}} = \vec{y}_i - \vec{y}_j + \vec{l}$, 
where $\vec{l}$ is 
a triplet of integers denoting the position of an image cell. One has
to point out that only the parameter $\beta$ enters into the equations
of motion. The second one $\Gamma$, enters through the choice of the
initial data: indeed, while the positions are
extracted from a uniform distribution, the velocities are taken from a
Maxwell distribution at temperature $T$. With this choice,
at the beginning of each simulation the system is out of
equilibrium: so there is a drift of the kinetic energy,
and the system reaches a different, random, temperature. In order to fix
the temperature to the desired value, we operate in this way: after
extracting the initial values, we let the system evolve until
equilibrium is reached, i.e. until the kinetic energy appears to
stabilize. We then generate new velocities again with a Maxwell
distribution at temperature $T$, and repeat the process until the kinetic
energy appears to be constant, close to the chosen value. 

Equations \eqref{newtonEquation} are integrated using a symplectic
splitting algorithm.  The inter particle forces are computed with the
aid of parallel calculators, using GPUs with up to 15 multiprocessors.
But even with such a device, we cannot afford to integrate the equations
of motion for small values of $\Gamma$, and we have to stop at
$\Gamma=0.01$. This for two problems which arise in the weakly coupled
regime.

The first problem concerns the integration step $h$. In fact, as the velocities 
are proportional to $\Gamma^{-1/2}$, to achieve a good energy conservation 
when short distance collisions occur, one has to use a very small time step. 
A step $h=10^{-3}$ is sufficiently small for the strongly coupled cases,
in which conservation of energy was always better than $0.05\%$.
The step had to be reduced up to $h=2.5\times10^{-5}$ for $\Gamma=0.01$. 
Curiously enough, it seems that also the value of $\beta$ has an 
influence on the choice of $h$. For example, if $\beta=10$, the value
$h=10^{-3}$ proved to be adequate even in the case $\Gamma=0.01$.
The error on conservation of energy is still below $0.05\%$ for 
$\Gamma=0.1$ but increases up to $0.5\%$ for $\Gamma=0.01$.

The second problem is that, working with periodic boundary conditions,
one should have the fundamental cell with side larger than the Debye
length $\lambda_D$, which in our units reads $\lambda_D =
\sqrt{1/\Gamma}$.  As in the rescaled variables the density has value
1, we have $L = N^{1/3}$, so that the requirement $L > \lambda_D$ in
our units takes the form of the constraint $N > \Gamma^{-3/2}$, which
is a very stringent condition in the weakly coupled regime $\Gamma \ll
1$.  Indeed, as the Coulomb force is a long range one, the
computational cost increases an $N^2$, i.e., increases as
$\Gamma^{-3}$. This means that the computational cost increases at
least a thousand times by decreasing $\Gamma$ by a factor
ten. Actually, as one must also decrease the value of the integration
step, the computational cost increases even more.  Now,  although in
the strongly coupled cases  a single particle would satisfy the
constraint, we actually used $480$ particles.  In the cases $\Gamma=
0.1$ and $0.01$ instead, for which the constraint gives $N>31$ and
$N>1000$ respectively, we took $N=896$ and $N=1024$ respectively. This
last figure is the maximum number of particles we can deal with.
Computations with this number of particles take months to be completed.

Our data are affected by some fluctuations, due to various factors;
mainly, we suppose, the very limited number of particles. In
paper~\cite{OttBonitz}, in which a model essentially equal to ours was
integrated, the authors report that a certain stability of the numerical
results is obtained using a number of particles equal to $N=8192$,
which is beyond our reach. So this work should be intended, also
numerically, as a preliminary one. Naturally, the main improvement would be
to be able to simulate a two-component model, which however is, at
the moment, far from our numerical capabilities.


\begin{thebibliography}{9}

\bibitem{OttBonitz} T.Ott, M.Bonitz, Phys. Rev. Lett. \textbf{107},
  135003 (2011).
\bibitem{Spitzer} L. Spitzer, Jr, \emph{Physics of Fully Ionized Gases} 
  (Dover Publications, New York, 1962).
\bibitem{Rosenbluth} C.L.Longmire, M.N.Rosenbluth, Phys. Rev.
  \textbf{103}, 507 (1956). 
\bibitem{CBMZG} A. Carati, F. Benfenati , A. Maiocchi , M. Zuin and
  L. Galgani, Chaos \textbf{24}, 013118 (2014).
\bibitem{CZMMMG} A. Carati, M. Zuin, A. Maiocchi, M. Marino,
  E. Martines, L. Galgani, Chaos \textbf{22}, 033124 (2012).
\bibitem{Wannier} G.~H.~Wannier, \emph{Statistical Physics} (Dover
  Publications, Inc., New York, 1987). 
\bibitem{Ranganathan} S.Ranganathan, R.E.Johnson, C.E.Woodward,
  Physics and Chemistry of Liquids, \textbf{14}, 123 (2003).
\bibitem{Gibbon} P. Gibbon, G. Sutmann, \emph{Quantum Simulation
  of Complex Many-Body Systems: from Theory to Algorithms},  
  in J.Grotendorst, D.Marx, A. Muramatsu eds, NIC Series \textbf{10},
  467-506 (2002). 



\end{thebibliography}
\end{document}